\documentclass[twocolumn]{aa}
\usepackage{graphicx}
\usepackage{txfonts}
\begin{document}

   \title{The environment of Low Surface Brightness Galaxies}

   \author{S.\,D. Rosenbaum%\inst{1}
          \and
          D.\,J. Bomans%\inst{1}          
          }

\offprints{S.\,D. Rosenbaum}

   \institute{Astronomisches Institut, Ruhr-Universit\"at Bochum (AIRUB),
              Universit\"atsstr. 150, 44780 Bochum, Germany\\
              \email{dominik.rosenbaum@astro.rub.de, bomans@astro.rub.de}}

   \date{Received 23 April 2004 / accepted 26 May 2004}

\abstract{
Using the Early Data Release of the Sloan Digital Sky Survey (SDSS) 
we investigated 
the clustering properties of Low Surface Brightness (LSB) galaxies 
in comparison to normal, High Surface Brightness (HSB) galaxies. 
We selected LSB galaxies and HSB galaxies with well measured redshifts 
from the SDSS data base and 
performed three-dimensional neighbour counting analysis within spheres 
of radii between 0.8\,Mpc and 8.0\,Mpc.  As a second analysis method we used 
an $N^{\rm th}$ neighbour analysis with $N$ varying from one to ten galaxies. 
Our results show significant differences between the galaxy densities of  
LSB galaxies and HSB galaxies on scales from 2 to 5\,Mpc. At scales larger 
than 5\,Mpc LSB and HSB galaxies share the same clustering 
properties.  In the pie-slice diagrams the LSB galaxies appear to favour 
the inner rims of filaments as defined by the HSB galaxies, with a couple 
of LSB galaxies even being located inside the voids.  
Our results support the idea of gas-rich LSB galaxies forming and developing 
in low density regions without many galaxy interactions and just now reaching 
the filaments of the large scale structure. 
}

   \maketitle
%
%________________________________________________________________

\section{Introduction}
The existence of gas-rich disk-galaxy-like LSB galaxies with central surface 
brightnesses of $\mu_B > 22.5$\,mag/arcsec$^2$ has been established over the 
last 15 years, whereas the formation and evolution scenarios that led to  
such a class of galaxies with a sparse stellar population are not well 
understood so far (e.g., Impey \& Bothun 1997). 
Although LSB galaxies have HI components with low surface densities 
(e.g., van der Hulst et al. 1993) systematically below the 
Kennicutt (1998) criterion for star formation, 
they can be regarded as gas-rich in general (e.g., Pickering et al. 1997).

The key in understanding LSB galaxies 
lies then in the answer to the question what prevented them from sufficient 
star formation. 
One explanation might be found in the differences in the 
spin parameter of the dark matter halo between LSB and HSB galaxies. 
Dalcanton et al. (1997) and Boissier et al. (2003) found 
some evidence that LSB galaxies could be disk galaxies with a larger spin 
parameter than HSB spirals. 
This would imply larger scale lengths for LSB disks and originate the 
observed low HI surface densities.    

Another way to reconstruct the evolutionary history and to understand 
the properties of these galaxies might be found in the nature of the small and 
large-scale environments in which LSB galaxies are embedded, since the lack 
of star formation can only occur if the galaxies were formed in low density 
regions.  
Only a low density scenario can 
warrant that neither tidal encounter with companions nor infall of massive 
gas clouds could have taken place and have triggered a sufficient star 
formation which would have gradually brightened the stellar disk.
Evidence for the stronger isolation of LSB galaxies in comparison to HSB 
galaxies was found before by Bothun et al. (1993) and Mo et al. (1994). 
A lack of nearby ($r \leq 0.5$\,Mpc) companions of LSB galaxies was detected 
by Zaritsky \& Lorrimer (1993). 

Today, with the availability of several substantial galaxy redshift surveys 
containing high quality redshifts, the possibility for intensive studies on 
the environmental galaxy densities of LSB galaxies is given.  

\section{Data Characteristics and Analysis}
The Early Data Release (EDR) of the Sloan Digital Sky Survey 
(SDSS, Stoughton et al. 2002) covers  around 460\,deg$^2$ of imaging data and 
54,000 spectra mainly from scans in two equatorial stripes, one in the 
southern and the other in the northern Galactic cap. 

We retrieved LSB candidates and a HSB comparison sample 
from the EDR using the SDSS Query Tool. 
The following 
parameters of each object, classified as a galaxy with spectroscopic data 
avaliable were downloaded: 
an object identifier, right ascension, 
declination, an azimuthally averaged radial surface brightness profile in the 
$g$- and $r$-band and the redshift of the object (for technical details to 
the SDSS parameters see chapter 4.10. in Stoughton et al. 2002).
The sample was limited to the two equatorial stripes of the EDR, a redshift 
of $0.02\leq z\leq0.1$ and in order to minimize the uncertainty of the redshift
 a z-confidence greater than 90$\%$ was demanded. 
A sample of 16123 galaxies was obtained.
 \begin{figure}\label{Pie}
\begin{center}
\begin{minipage}[t]{8.05cm}
\flushleft
   \includegraphics[width=8.0cm]{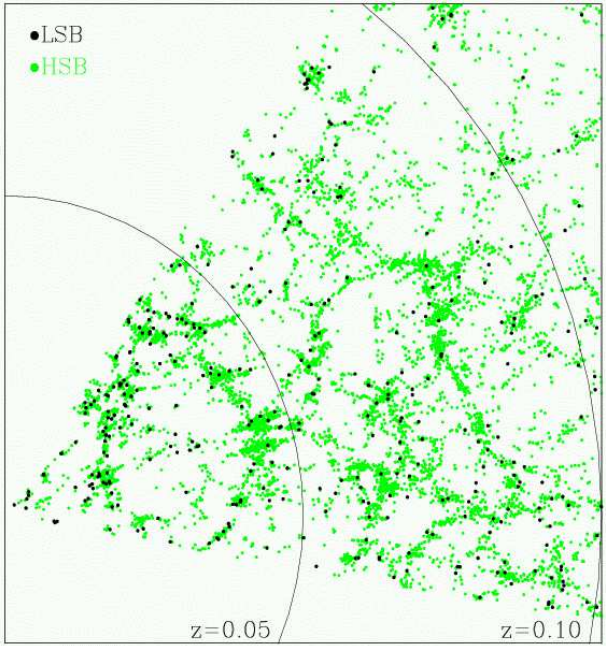}
   \caption{One of the two analysed pie slices. The right ascension range
$354\degr\leq \alpha \leq\,53\degr$ 
containing $\sim$400 LSB galaxies and $\sim$8000 HSB galaxies is plotted.}

 \end{minipage}
\end{center}
 \end{figure}

Equipped with this data set we were able to 
distinguish between LSB galaxies with a central surface brightness 
$\mu_B>22.5$\,mag/arcsec$^2$ and HSB galaxies (if 
$\mu_B\leq 22.5$\,mag/arcsec$^2$). 
The central surface brightness $\mu_B$ was calculated for each galaxy 
from the central annulus ($0.23''$) of the radial surface 
brightness profile in the filters $g$ and $r$ using an equation following 
Smith et al. (2002).
At the faint end the surface brightness distribution was cut off at 
$\mu_B=25$\,mag/arcsec$^2$. 
We did not apply any selection criterion based on the apparent size of the 
galaxies since the intrinsic selection criterion of the SDSS  
spectroscopic sample limits the angular diameter of the galaxies on which 
spectroscopy was applied to a value of 5\arcsec or larger.
With these selection criteria a sample of 804 LSB galaxies from both 
equatorial stripes of the EDR with a redshift of $z \leq 0.1$ was obtained. 
In this data set 15319 Galaxies remained as HSB galaxies.

For further analysis of the LSB environment and density distribution, two 
different methods were used. 
First, the environmental density for each sample LSB and HSB galaxy was 
measured by counting the number of galaxies within a sphere around the 
scrutinised galaxy. 
Second, the distance to the $N^{\rm th}$ (with $N$ 
between 1 and 10) nearest neighbour of each galaxy
was determined and used as a measure of the local density. The separation
between LSB and HSB galaxies was performed after statistical environmental
analysis in both methods.

\begin{subsection}{Neighbour Counting within Spheres} 
The program for neighbour counting was fed with the catalog file of the  
sample containing both LSB and HSB galaxies downloaded as described before.
First the code converted the
redshift, right ascension and declination of all galaxies from our sample into 
the 3 dimensional spatial distribution of galaxies. 
For that a Hubble constant of 
$H_0=71$\,km/s (Bennet et al. 2003) was used.
The purpose of the algorithm was to deliver the number of neighbours within a 
sphere of a certain radius for every single galaxy. 
The radius of the sphere was a fixed parameter during each run.
In several runs the radius of the sphere was varied
between $r=0.8$\,Mpc and  $r=8.0$\,Mpc in steps of 0.6\,Mpc. In order to avoid 
biasing of the sample due to boundary effects, the program limited the sample 
to galaxies with distances to the boundary of the catalog volume larger 
than the radius of the sphere. Due to this edge correction, the inspected 
volume varies depending on the radius $r$.  
For this analysis both equatorial stripes were used. 
After this neighbouring analysis the resulting statistical distribution 
was divided into LSB and HSB subsamples using the central surface brightness 
criterion $\mu_B > 22.5$\,mag/arcsec$^2$ for low surface brightness galaxies 
as described above.
\begin{figure*}\label{Sphere}
   \centering
   \includegraphics[width=7.5cm]{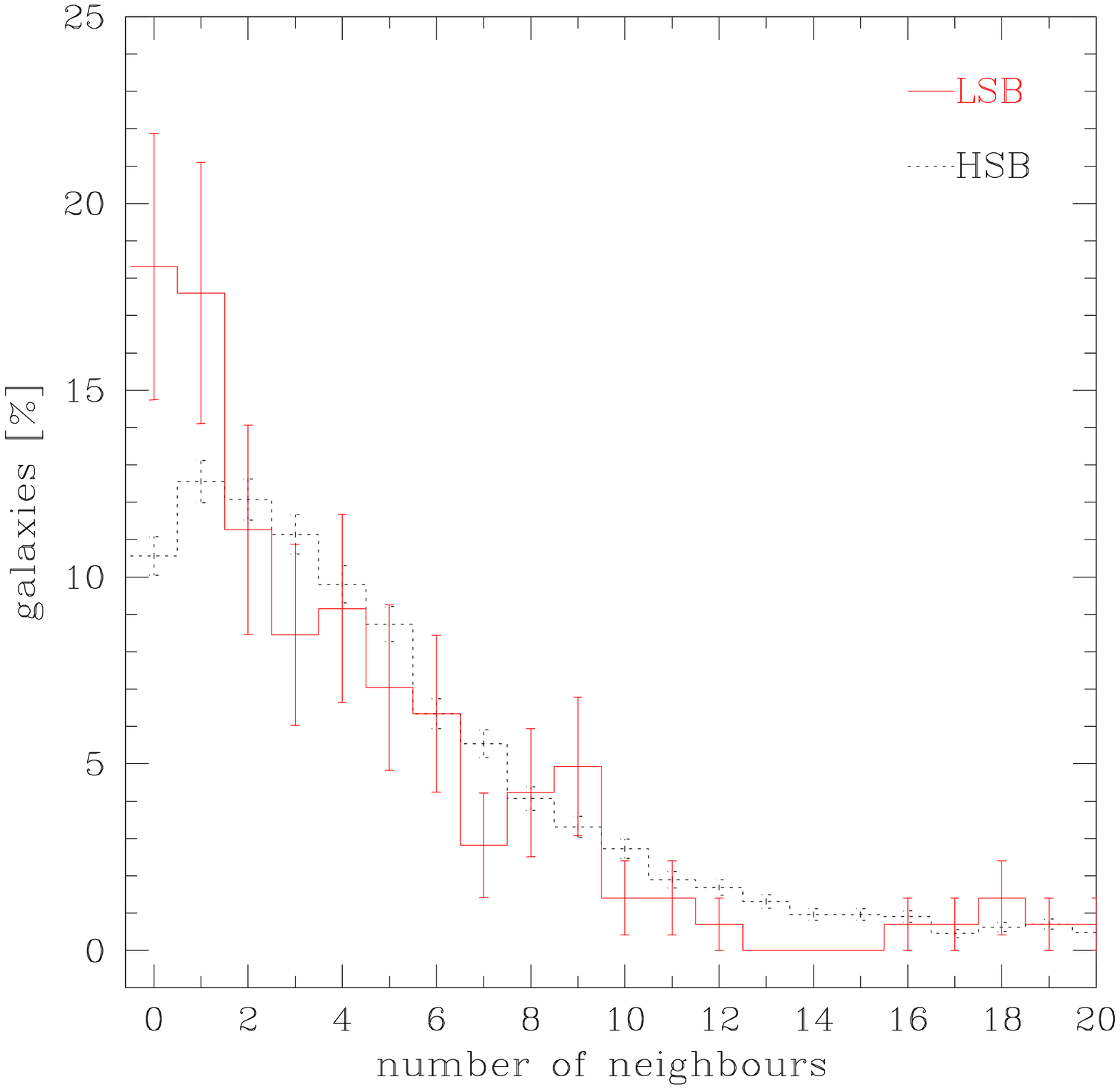}
   \includegraphics[width=7.5cm]{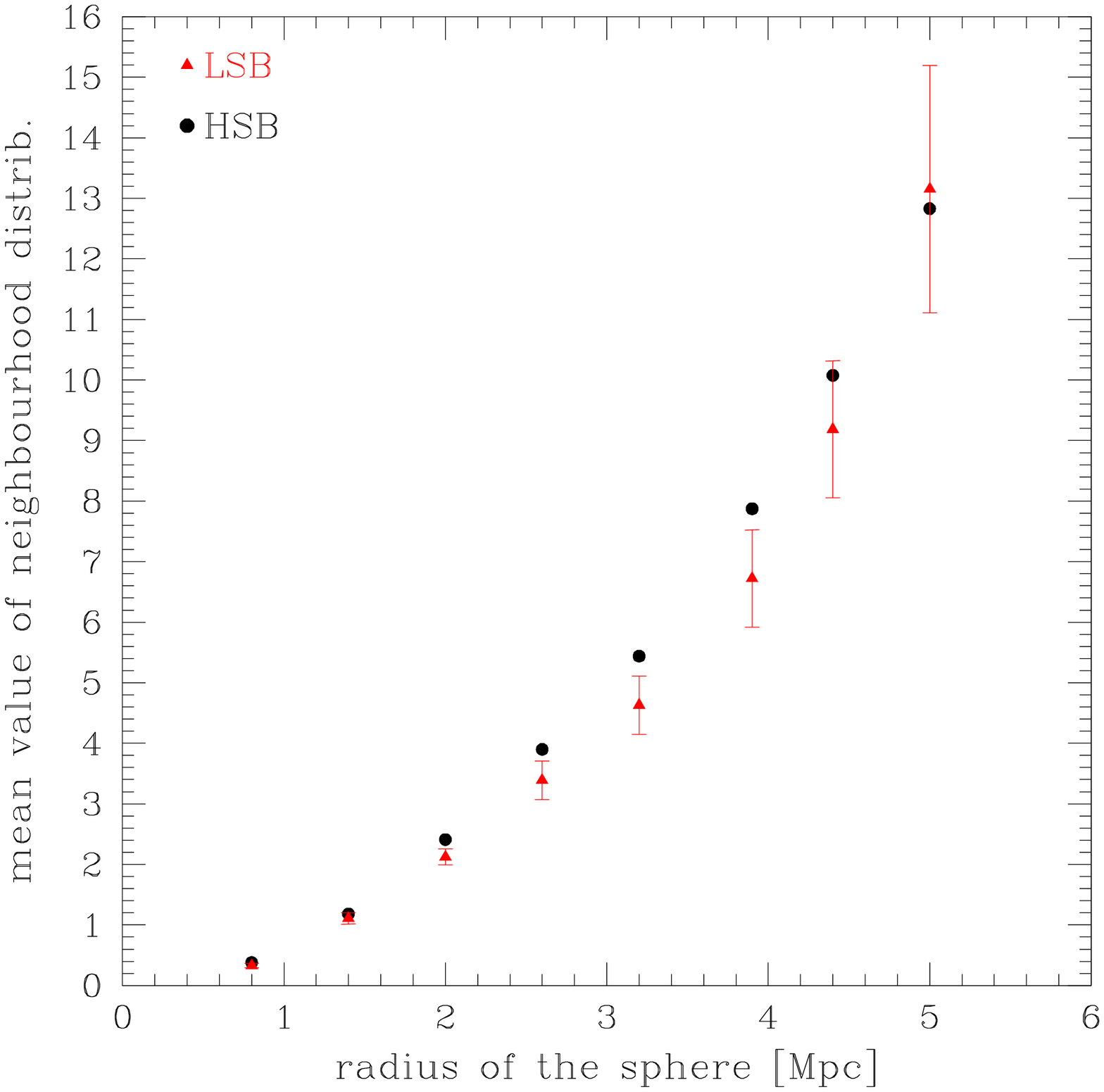}

   \caption{Left panel shows the distribution of the number of neighbours 
   within a sphere with a radius of $r=3.2$\,Mpc for
   LSB (solid line) and HSB (dashed line) galaxies. Right panel contains the
   mean values of the neighbourhood distributions for LSB (triangle) and HSB
   galaxies (dots) at several sphere radii.}%
    \end{figure*}
\end{subsection}
\begin{subsection}{$N^{\rm th}$ Neighbour Analysis}
The second analysis using the $N^{\rm th}$ nearest  
neighbour method was limited to the first equatorial stripe which corresponds
to the right ascension area of 
$354\degr \leq \alpha \leq 53\degr$. 
The right ascension, declination and redshift informations 
from the sample catalog were again translated to a 3-dimensional grid. 
The distances between the galaxies were calculated and the $N^{\rm th}$ 
lowest distance to the neighbouring galaxies was assigned to each member 
of the sample as a measure of local galaxy density. 
Then an edge correction was applied by
rejecting all galaxies with distances to the $N^{\rm th}$ 
neighbour larger than to
the edge of the sample catalog. At the end, the 
resulting sample was divided into LSB and HSB subsamples.
This kind of analysis was repeated for several values of $N$ varying 
from 1 to 10.
\end{subsection}

\section{Results}
In Fig. 1 the distribution of LSB galaxies in 
comparison to HSB galaxies is shown in a so called pie slice diagram, where 
the right ascension and the redshift of the galaxies are 
displayed in a polar plot (the declination range is projected onto the plane). 
The pie slice diagram shows 
that LSB galaxies are located in the filaments of the 
Large Scale Structure (LSS) traced by the 
distribution of HSB galaxies. Further investigations of the plot lead 
to the impression that LSB galaxies show the tendency to be located more 
often at the edges of these filaments than in the center and some LSB 
galaxies are even found in void regions.
 
In order to amplify this impression the statistical environmental study 
as described before was performed.
The results of the analysis method using number counting within spheres were 
plotted in a diagram for each sphere radius 
showing the  neighbouring statistics for LSB and HSB galaxies at several 
scales.
Figure 2 shows as an example the distribution for the 
number of neighbours within a sphere of the radius
$r=3.2$\,Mpc (left panel) around each HSB and LSB galaxy. 
The mean values of neighbours for LSB and HSB galaxies were calculated for 
every applied sphere radius and are displayed in the right panel as a 
function of the corresponding radius. 
The error bars indicate the statistical error for LSB mean values 
(error bars for the HSB distribution correspond approximately to the size of 
the dots). 
In Fig. 3 the results of this analysis for the runs 
with $N=1$ and $N=5$ of the distributions of LSB and HSB galaxies are 
displayed. In order to increase the low number statistics of the LSB sample 
a cumulative delineation is chosen.

On lower scales (with $r$\,=\,2.0\,Mpc and below, Fig. 2 right panel and 
Fig. 3) 
the environmental statistics of the LSB galaxy distribution follows the 
distribution generated by the HSB galaxies within error margins as well. 
However, 
all LSB mean values lie systematically (but partially not significantly) below 
the corresponding HSB value. 

The left panel of Fig. 3 shows that 
LSB galaxies seem to participate in pairs on the same scales
as pure HSB pairs (represented by the HSB curve in the diagram). 
We want to point out that in our analysis the number of LSB-HSB 
pairs, which are counted both as HSB and as LSB pairs, is negligibly small 
compared to the quantity of HSB-HSB pairs. This is due 
to the fact that the amount of HSB galaxies exceeds the number of LSB galaxies
by a factor of 20 (LSB-LSB pairs do not appear in our sample).

Studies of the environment at larger scales 
($2.0\leq r \leq 5.0$\,Mpc) led to different results.
While at small scales ($r\leq 2.0$\,Mpc) the percentage of LSB galaxies with no
or one neighbour is nearly the same as for HSB galaxies, 
this fact does not apply at 
intermediate scales ($2\leq r\leq 5$\,Mpc, Fig. 2, left panel).
This result is reproduced by the dependency of the mean values on 
the sphere radius  (right panel, Fig. 2). 
On scales between 2 and 4\,Mpc the mean values are significantly 
lower for the LSB galaxies than for the HSB sample.
In order to proof the significance of the statistics a Kolmogorov-Smirnov 
(KS) test was performed. For $2.6\leq r \leq 4$\,Mpc the KS test rejects 
the null hypothesis that the distributions for LSB and HSB neighbours 
are the same at a confidence level greater than $92\%$ with a maximum value 
of $99.3\%$ at a radius of 2.6\,Mpc. 

These results are also consistent with our $N^{\rm th}$ neighbour analysis. 
The 5$^{\rm th}$ nearest neighbour analysis (Fig. 3, right panel) shows a gap
between both cumulative distributions on scales between 3.2 and 5.0\,Mpc.
This means that significantly fewer LSB galaxies exist with a
distance to the 5$^{\rm th}$ neighbour on these scales than HSB galaxies. 
This phenomenon reappears in the distributions of 3$^{\rm rd}$ and 
4$^{\rm th}$ neighbour analysis but shifted slightly to lower scales. 
For the 4$^{\rm th}$ neighbour analysis this gap is 
located within the interval of 3.0 to 4.8\,Mpc  and for 3$^{\rm rd}$ 
neighbour studies it is found between 2.8 and 4.4\,Mpc. The neighbourhood 
investigations for $N=4$ and $N=5$ are important, because 
they correspond to the typical number of compact group members. 
For $N\geq$ 6 the cumulative distribution of the LSB galaxies 
follows the HSB distribution.
        
\section{Discussion and Conclusions}

Figure 1 shows that the spatial distribution of LSB galaxies 
follows in general the LSS defined by HSB galaxies which is in  good 
agreement with the results from investigations on LSB galaxies in the 
Century Survey (Brown et al. 2001). 
However, as mentioned before, there are some extremely isolated LSB galaxies 
located in voids of the LSS. 
The statistical results show that the isolation of 
LSB galaxies takes place on intermediate scales beyond the  
size of compact groups but in the range of the size of 
large groups and LSS filaments as well.
For smaller ($r\leq 2$\,Mpc) and 
larger scales ($r\geq 5$\,Mpc) no significant differences in the statistical 
environments could be found in this study. 

\begin{figure*}\label{NNeighbour}
   \centering
   \includegraphics[width=7.5cm]{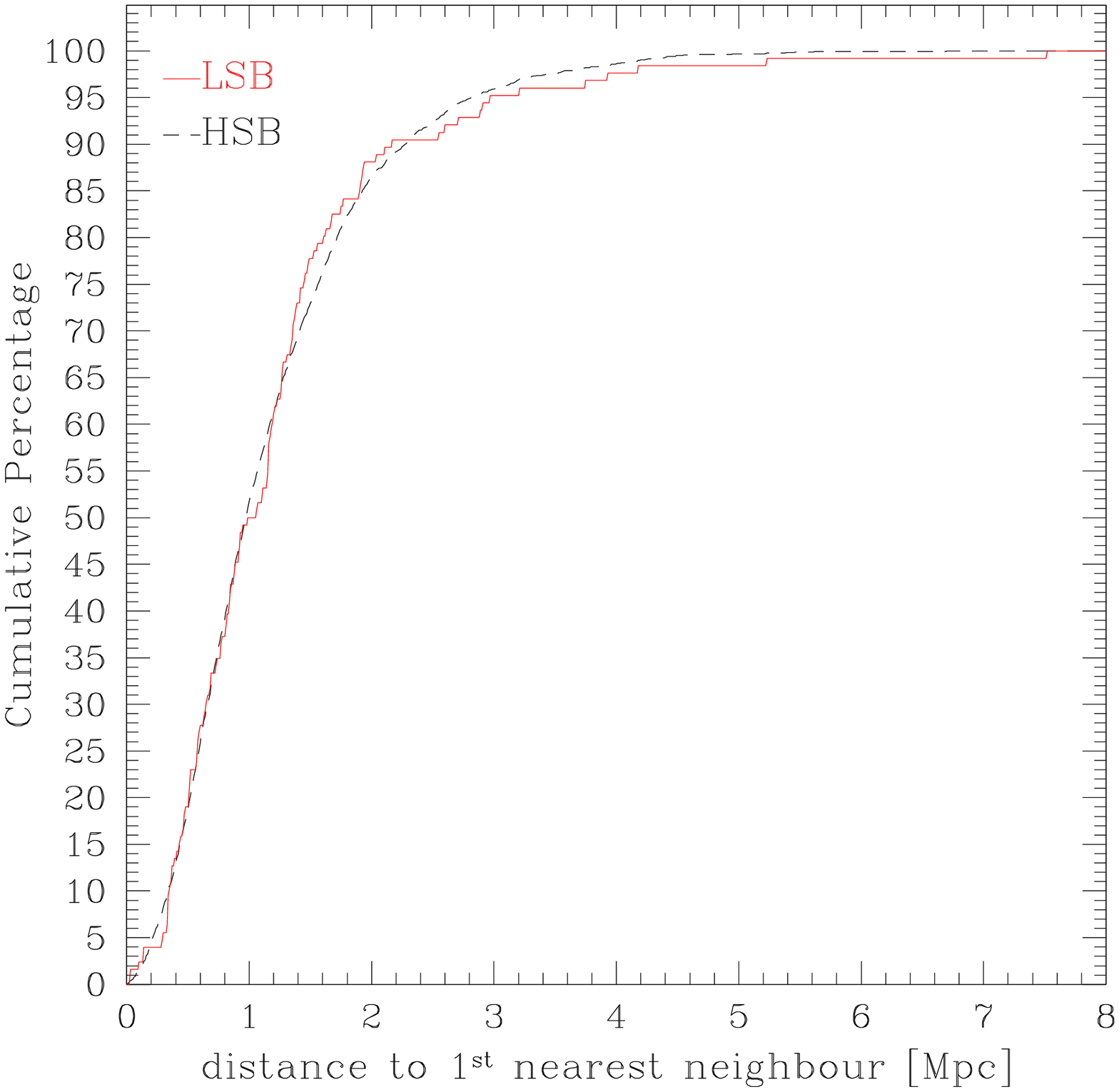}
   \includegraphics[width=7.5cm]{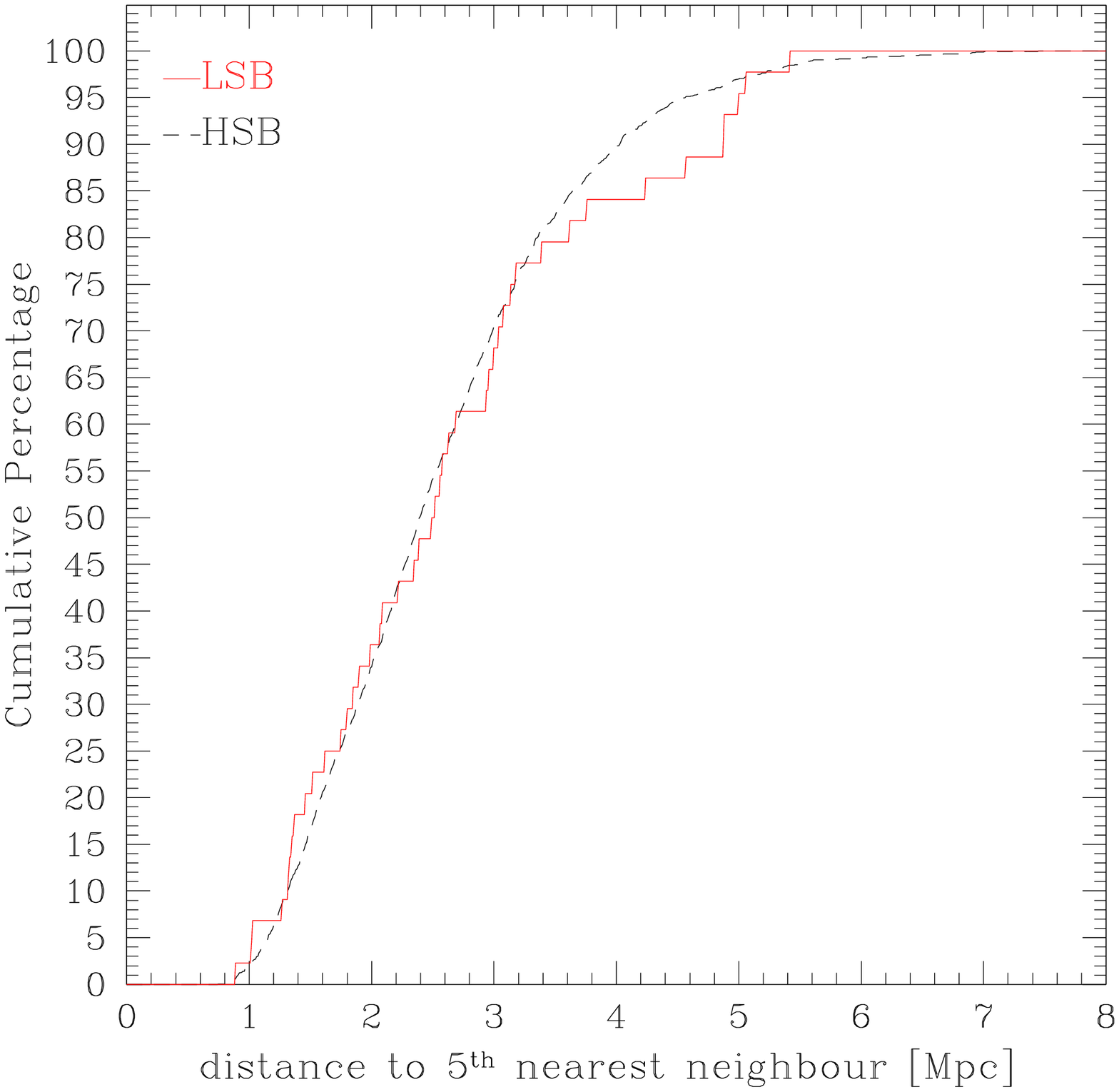}
   \caption{Left panel shows the cumulative distribution of the distances
   to the first neighbour for LSB (solid line) and HSB (dashed line)
   galaxies. On the right panel the same distribution for the fifth nearest 
    neighbour analysis is plotted.}
    \end{figure*}

Our results contradict the results of Bothun et al. (1993), 
who found that the neighbouring distribution of 
LSB galaxies differs only on scales below 2\,Mpc from the HSB distribution.
Their statistics might be biased towards smaller scales by 
projection effects due to their two-dimensional treatment of the problem, 
which might have lead them to see the differences at smaller scales.
However we cannot rule out the existence of differences between the LSB and 
HSB distributions on scales with $r\leq 2$\,Mpc since these 
scales  
cannot be resolved statistically in our study due to the small numbers of
neighbours at these scales in our sample.
We only see a slight tendency that LSB galaxies may have less 
neighbours on small scales. Clearly this has to be investigated using 
the much larger datasets of the SDSS Data Releases 1 and 2 
(Abazajian et al. 2003).
On scales around 5\,Mpc and beyond LSB galaxies trace the identical 
structure as HSB galaxies consistent 
with the results from  Bothun et al. (1993).

All our results fit well into the following formation scenario, which 
was proposed by e.g. Bothun et al. (1997):
galaxy formation takes place due to an initial Gaussian spectrum 
of density perturbations with much more low-density fluctuations than high 
density ones. 
Many of these low-density perturbations are lost due to the assimilation
or disruption during the evolutionary process of galaxy formation but a 
substantial percentage of the fluctuations survives and is expected to form 
LSB galaxies.
Further on one can assume that the spatial distribution of the initial 
density contrast consists of small scale fluctuations superimposed on
large-scale peaks and valleys.  
Small-scale peaks lead to galaxy formation, whereas the large-scale maxima 
induce cluster and wall formation of the LSS.           

Based on our results presented here, we propose that the
galaxies formed in the large-scale valleys may develop to LSB galaxies 
due to their isolated environments whereas HSB galaxies formed mainly on the 
large-scale peaks.
The isolation of LSB galaxies on intermediate and small scales must have 
effected their evolution since tidal encounters acting as triggers for star 
formation would have been rarer in these LSB galaxies than for HSB galaxies. 
Our results give strong evidence for this scenario, since the observed 
isolation of LSB galaxies takes place on scales below 5\,Mpc, which is 
exactly the typical size of LSS filaments (e.g., White et al. 1987, 
Doroshkevich et al. 1997). 

Hence, we conclude that 
LSB galaxies were formed in the voids of the LSS and that most of them 
have migrated to the edges of the filaments due to gravitational infall, but 
some of them still remain in the voids where they have formed in.

\begin{acknowledgements}{}
We are grateful to Ralf-J\"urgen Dettmar, Zita Banhidi, Anja von der Linden, 
Lutz Haberzettl, and Steffen Mieske for 
proof-reeding and precious incitations to this paper. We thank the referee
Karl Rakos for his  comments.
This work was supported financially by the GRK\,787 ``Galaxy Groups as 
Laboratories for Baryonic and Dark Matter'', and DFG project BO1642/4-1.
Funding for the creation and distribution of the SDSS Archive has been 
provided by the Alfred P. Sloan Foundation, the Participating Institutions, 
the NASA, the National Science Foundation, the US Department of Energy, the 
Japanese Monbukagakusho, and the Max Planck Society. 
%(www.sdss.org). 
\end{acknowledgements}

\end{document}